# Distinguishing dynamical features of water inside protein hydration layer: Distribution reveals what is hidden behind the average

Saumyak Mukherjee[†a], Sayantan Mondal[†b] and Biman Bagchi[*]

[†*]Solid State and Structural Chemistry Unit

Indian Institute of Science, Bengaluru-560012, India

## Abstract

*Since the pioneering works of Pethig, Grant and Wüthrich on protein hydration layer, many studies have been devoted to find out if there are any "general and universal" characteristic features that can distinguish water molecules inside the protein hydration layer from bulk. Given that the surface itself varies from protein to protein, and that each surface facing the water is heterogeneous, search for universal features has been elusive. Here, we perform atomistic molecular dynamics simulation in order to propose and demonstrate that such defining characteristics can emerge if we look not at average properties but the distribution of relaxation times. We present results of calculations of distributions of residence times and rotational relaxation times for four different protein-water systems, and compare them with the same quantities in the bulk. The distributions in the hydration layer is unusually broad and log-normal in nature, due to the simultaneous presence of peptide backbones that form weak hydrogen bonds, hydrophobic amino acid side chains that form no hydrogen bond and charged polar groups that form strong hydrogen bond with the surrounding water molecules. The broad distribution is responsible for the non-exponential dielectric response and also agrees with large specific heat of the hydration water. Our calculations reveal that while the average time constant is just about 2-3 times larger than that of bulk water, it provides a poor representation of the real behaviour. In particular, the average leads to the erroneous conclusion that water in the hydration layer is bulk-like. However, the observed and calculated lower value of static dielectric constant of hydration layer remained difficult to reconcile with the broad distribution observed in dynamical properties. We offer a plausible explanation of these unique properties.*

---

[*] *Corresponding author* (profbiman@gmail.com)

[†] S. Mukherjee and S. Mondal contributed equally

# 1. INTRODUCTION

A layer of water that surrounds every protein molecule in aqueous solution plays a central role in the structure, dynamics and function of the protein[1-22]. An early estimation of the width of the hydration layer came from the rotational correlation time obtained by NMR (and later by dielectric relaxation) measurements of the protein in aqueous solution. The measured time constant was found to be elongated due to the interaction with the surrounding water molecules. Use of Debye-Stokes-Einstein relation to reproduce the orientational correlation time demonstrates the need for an addition of ~3Å to the crystallographic radius of the protein[23]. This 3Å seemed correct to accommodate one layer of water. This tentative agreement served to foster the view that a protein in aqueous solution is surrounded by a nearly rigid layer of water molecules (the iceberg model). The landmark work of Wüthrich dispelled this idea to some extent by suggesting that the residence time of water molecule in the layer should be less than ~300 ps[17, 18].

Even earlier than the reported NMR experiments, Pethig and others studied aqueous protein solutions by using dielectric spectroscopy[14-16, 24]. They essentially discovered three components that were considered universal by many, including Mashimo[25, 26] who carried out extensive studies in the late 1980s. The three components consist of (i) one bulk water-like around 10 ps, (ii) one at 10 ns or so, attributed to protein rotation and the third (iii) at around 40 ps. The last one was unexpected and was termed '*delta-dispersion*'. This was attributed to protein hydration layer (PHL).

Much later, the problem was re-visited by employing improved NMR techniques[27-29], time dependent fluorescence Stokes shift (TDFSS) studies[12, 30, 31] and also computer simulation studies[2, 32-34]. New NMR experiments all but rule out existence of any slow component[28, 35]. *The average time obtained was only 2-3 times slower than that of the bulk value*. On the other hand, recent TDFSS experiments consistently produced time component that were more than one order of magnitude slower than that in the bulk[13, 30, 36-41]. Let us first focus on results obtained by NMR experiments. By the very nature of the experimental technique, NMR provides only an average value, that is, average over all the water molecules in the system.[29] That is, both in the surface and away. One can use NOE or spin exchange technique to obtain region specific result but NOE has low time resolution. MHRD on the other hand has little or no spatial resolution[29, 35]. The inability of NMR to provide either spatial or temporal resolution makes it hard to apply to draw any definite conclusion. TDFSS on the other hand reported existence of several slow components, ranging from tens of ps to hundreds of ps[30, 33]. However, interpretation of the origin of slow components remains controversial to-date. Initial experiments by Bhattacharyya and co-workers revealed the existence of time scales ranging from a few ps to even a few ns[38]. However, these experiments had limited time resolution so missed much of the ultrafast response.

Zewail and co-workers carried out experiments on *Subtilisin Carlsberg* and sweet protein *Monellin* using exposed amino acid residues (tryptophan) as the natural probe[30]. Because of 160 fs time resolution used in these experiments, they missed both the ultrafast and the slow components but obtained the intermediate timescales. Importantly, they compared their TDFSS results on protein hydration layer with tryptophan in the bulk. Zewail's experiments find a slow component of 38 ps for *Subtilisin Carlsberg* and 16 ps for sweet protein *Monellin* which are absent in bulk water solvation.

Computer simulations, however, have provided mixed results. If one uses single particle rotation and probe the second rank spherical harmonic (as in anisotropic depolarization experiments) then one finds a result in good agreement with NMR, that is, a relaxation time ~2-3 times slower than the bulk. On the contrary, if one studies dielectric relaxation or the total moment-moment time correlation function of the first layer[42], then one obtains a multi-exponential decay with the slowest time that is again an order of magnitude slower than the bulk[42]. It is perhaps expected that different experimental techniques would lead to different results and different conclusions. For example, it was pointed out by Hubbard and Wolynes[43], and also by Ravichandran and Bagchi[44] that dipolar interaction makes the rank ($l$) dependence of orientational relaxation non-trivial. The Debye *l(l+1)* dependence of the rate of relaxation might not hold[31].

In an interesting study, Ali and Singer pointed out that the amino acid side chains can play an important role in slowing down the solvation dynamics of a probe[32]. When they quenched the motion of the side chains, relaxation became faster. One could imagine that this is a trivial consequence of the removing the slow energy component from the side chain charged groups, but later study showed that the situation was not that simple. In some cases, the relaxation became slower when side chain motion was quenched[33]. Therefore, a more detailed study is needed in a microscopic scale.

The main results of the present work are as follows. (i) Distributions of calculated residence times and rotational relaxation times in the hydration layer for four different protein-water systems are unusually broad. We attribute this to the simultaneous presence of peptide backbones that form weak hydrogen bonds, hydrophobic amino acid side chains that form no hydrogen bond and charged polar groups that form strong hydrogen bond with the surrounding water molecules. (ii) Importantly, this unusually broad distribution is responsible for the non-exponential relaxations. (iii) While the average time constant is just about 2-3 times larger than that of bulk water, it is seen to provide a poor representation of the real behaviour. In particular, the average leads to the erroneous conclusion that water in the hydration layer is bulk-like. (iv) The much lower value of the static dielectric constant of hydration layer remained difficult to reconcile with the broad distribution observed in the dynamical properties. We offer a plausible explanation of these unique properties.

We also discuss the relationship of our result of wide distribution of relaxation times with the experiments, like NMR, 2D-IR and time dependent fluorescence Stokes shift. We discuss how these different experiments preferentially probe different aspects of this distribution, and can thus lead to different results, leading to certain unnecessary confusion and controversy.

The organization of the rest of the paper is as follows. First, we try to show how the PHL is different from bulk solvent with respect to (i) first and second rank orientational correlation time constants ($\tau_1$ and $\tau_2$) of the hydration layer water molecules which account for rotational diffusion and (ii) 'Translation time' distribution of water molecules in the PHL and quantification using heaviside step function formalism that accounts for the translational diffusion and. Second, we calculate two equilibrium properties of the successive hydration layers, namely effective dielectric constant ($\varepsilon_{eff}$) and specific heat ($C_v^{eff}$) in comparison to that of the bulk water. Third, we show how the dynamics of solvation of a spherical virtual probe changes as it resides at various sites inside PHL. The conclusions are drawn based on these results obtained for four protein-water systems; namely antimicrobial protein *Lysozyme* (PDB ID: 1AKI), oxygen storage and transport protein *Myoglobin* (PDB ID: 3E5O), immunoglobulin binding *Protein-G* (PDB ID: 2GB1) and sweet protein *Monellin* (PDB ID: 2O9U) in order to distinctly characterise and draw general remarks on the hydration layer and its uniqueness. The four proteins are chosen because of their diverse structure, function and helix-sheet ratio (see **Figure 1** for details).

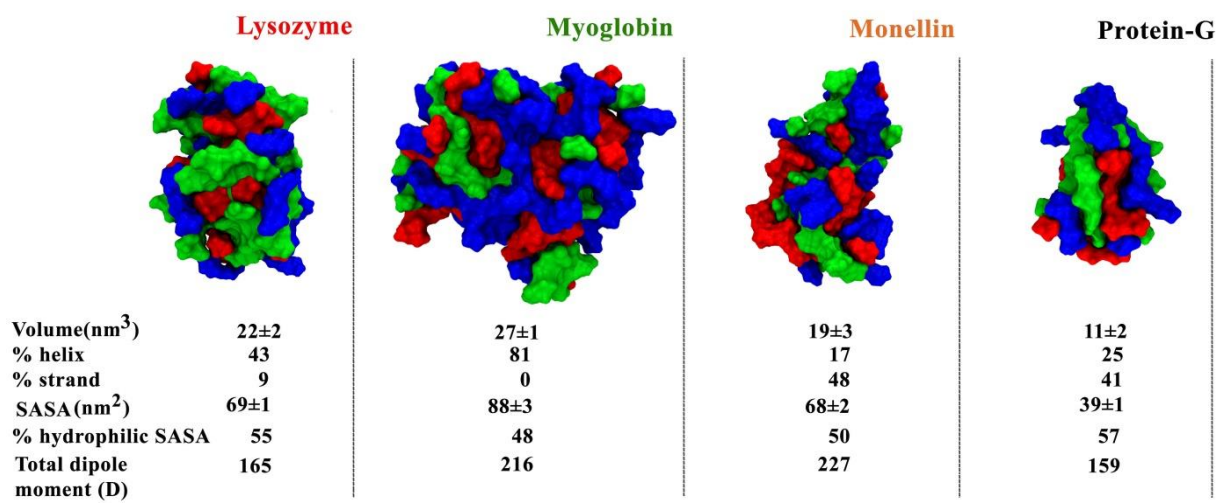

**Figure 1.** Surface representations of four model protein systems along with some crucial parameters. Hydrophobic residues are shown in blue; polar and uncharged residues are shown in green and charged residues are in blue. The percentage of different secondary structures are obtained using Stride package[45]. Average volume and SASA have been calculated using Gromacs[46] from 20 ns trajectories. The figures have been prepared using VMD[47].

## 2. SYSTEM AND SIMULATION DETAILS

Atomistic molecular dynamics simulations are performed using GROMACS[46] package (v5.0.7). We have prepared the system in accordance with experimental concentration (~2-3 mM). Initial configurations of the proteins have been taken from crystal structures available in Protein Data Bank. We have used OPLS-AA force field[48] and extended point charge (SPC/E) water model. Periodic boundary conditions were implemented using cubic boxes of sides ~9-10 nm filled with ~23,000-26,000 water molecules depending on the size of the protein. The total system was energy minimised using steepest descent algorithm followed by conjugate gradient method. Thereafter the system was subjected to simulated annealing[49] in order to heat it up from 300K to 320K and again cool it down to 300K in order to unbias the system and help it to get out of a local minima (if any). The solvent was equilibrated for 10 ns at constant temperature (300 K) and pressure (1 bar) (NPT) by restraining the positions of the protein atoms followed by NPT equilibration for another 10 ns without position restrain. The final production runs were carried out at a constant temperature (T=300K) (NVT) for 30 ns. Analyses were peformed on the last 25 ns of the trajectories to get rid of effects of barostat. The equations of motions were integrated using leap-frog integrator with an MD time step of 1 fs. All reported data are averaged over three MD trajectories starting from entirely different configuration of the system. We have used modified Berendsen thermostat[50] ($\tau_T$ = 0.1 ps) and Parrinello-Rahman barostat[51] ($\tau_P$ = 2.0 ps) to keep the temperature and pressure constant respectively. The cut-off radius for neighbour searching and non-bonded interactions was taken to be 10 Å and all the bonds were constrained using the LINCS[52] algorithm. For the calculation of electrostatic interactions, Particle Mesh Ewald (PME)[53] was used with FFT grid spacing of 1.6 Å.

## 3. RESULTS AND DISCUSSIONS

### 3.1 Distribution of rotational time constants

*One of the most interesting and somewhat unexpected outcomes of the present study is the observation of a broad distribution of relevant relaxation times obtained from rotational relaxation and translational diffusion of water molecules.* In **Figures 2** and **3** we show such distribution of relaxation times obtained for time correlation functions of several different dynamical quantities. Note the completely different nature of distribution compared to that of the bulk.

In order to characterise the distinctiveness in terms of rotation of one O—H bond of water molecules, we calculate the first and second rank orientational correlation [**Equations (1)** and **(2)**] for those water molecules which reside more than 100 ps inside the hydration layer and are monitored till they leave the PHL in order to obtain a good statistical averaging. We define a particular water molecule inside hydration layer only when it is within 1 nm of its nearest protein atom. For bulk solvent the distribution is calculated for ~4000 water molecules averaged over a 10 ns trajectory.

$$C_1(t) = \langle P_1(\hat{\mu}_0 . \hat{\mu}_t) \rangle; \text{ where } P_1(x) = x \qquad (1)$$

$$C_2(t) = \langle P_2(\hat{\mu}_0 . \hat{\mu}_t) \rangle; \text{ where } P_2(x) = \tfrac{1}{2}(3x^2 - 1) \qquad (2)$$

Here, $P_1$ and $P_2$ are respectively the first and second rank Legendre polynomials and $\mu_t$ are the unit vectors along any one O—H bond vector at time 't'. The thus obtained rotational time correlation functions for each individual water molecules are fitted to a multi-exponential function and the time constants are obtained by integrating the area under each curve. The distributions (histogram) of those time constants are also broad and log-normal in naure, with a long tail extending up to a few hundred ps (**Figure 2** and **Figure 3**). The averaged time correlation functions (i.e., averaged over all the water molecules considered) for each of the proteins are shown in **Figure 2** and **Figure 3** (inset) and the fitting parameters are noted down in **Table 1** and **Table 2**.

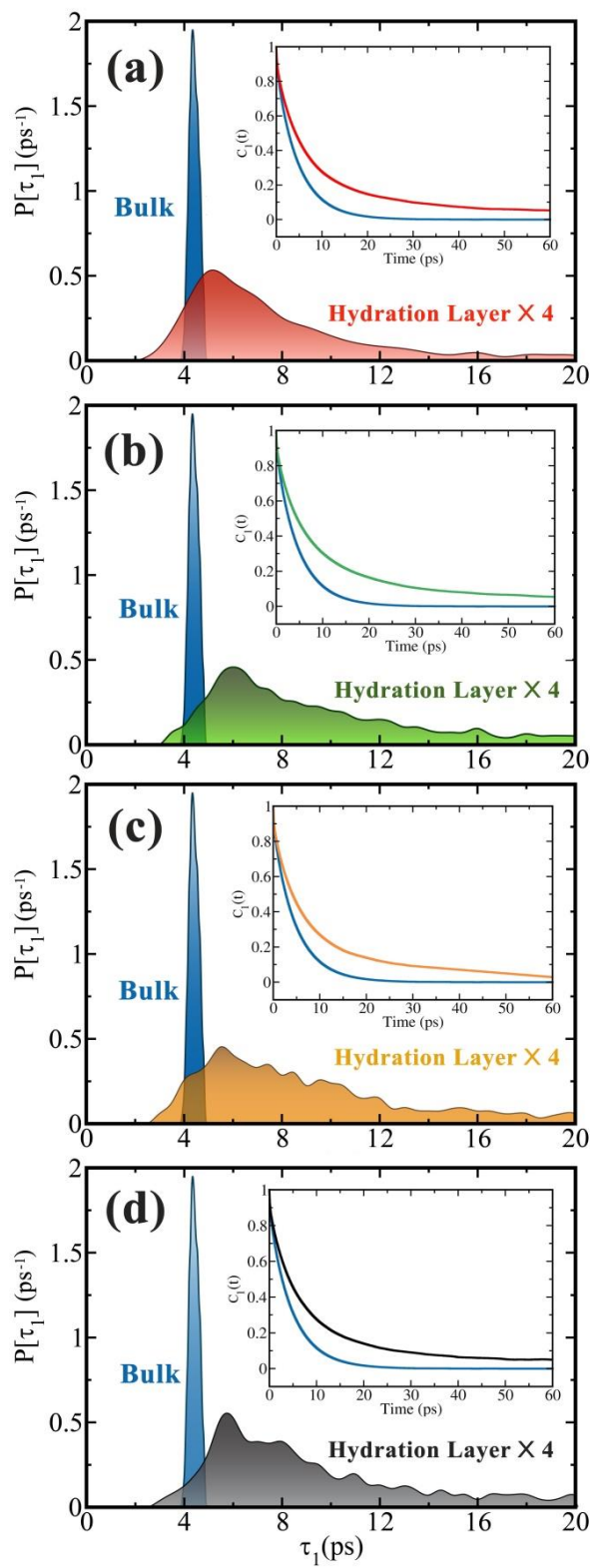

**Figure 2.** Distribution of first rank rotational time constants of water molecules inside protein hydration layer for four proteins and bulk (blue). In the insets the normalised and averaged rotational time correlation function is shown using same colour codes. (a) Lysozyme (b) Myoglobin (c) Monellin and (d) Protein-G.

**Table 1. Multi-exponential fitting parameters of the averaged and normalised first rank rotational relaxation of hydration layer water molecules and that of bulk. The slowest of the timescales (noted down in bold) was absent in bulk solvent.**

|  | $a_1$ | $\tau_1$(ps) | $a_2$ | $\tau_2$(ps) | $a_3$ | $\tau_3$(ps) | $<\tau>$(ps) | Average retardation |
|---|---|---|---|---|---|---|---|---|
| **Lysozyme** | 0.13 | 0.21 | 0.66 | 5.63 | **0.21** | **38.6** | 11.85 | 2.76 |
| **Myoglobin** | 0.16 | 0.34 | 0.63 | 6.38 | **0.21** | **40.3** | 12.53 | 2.92 |
| **Protein-G** | 0.14 | 0.29 | 0.68 | 6.14 | **0.18** | **41.4** | 11.66 | 2.72 |
| **Monellin** | 0.11 | 0.15 | 0.63 | 5.06 | **0.26** | **28.1** | 10.51 | 2.45 |
| **Bulk Water** | 0.13 | 0.21 | 0.87 | 4.93 | --- | --- | 4.29 | 1.00 |

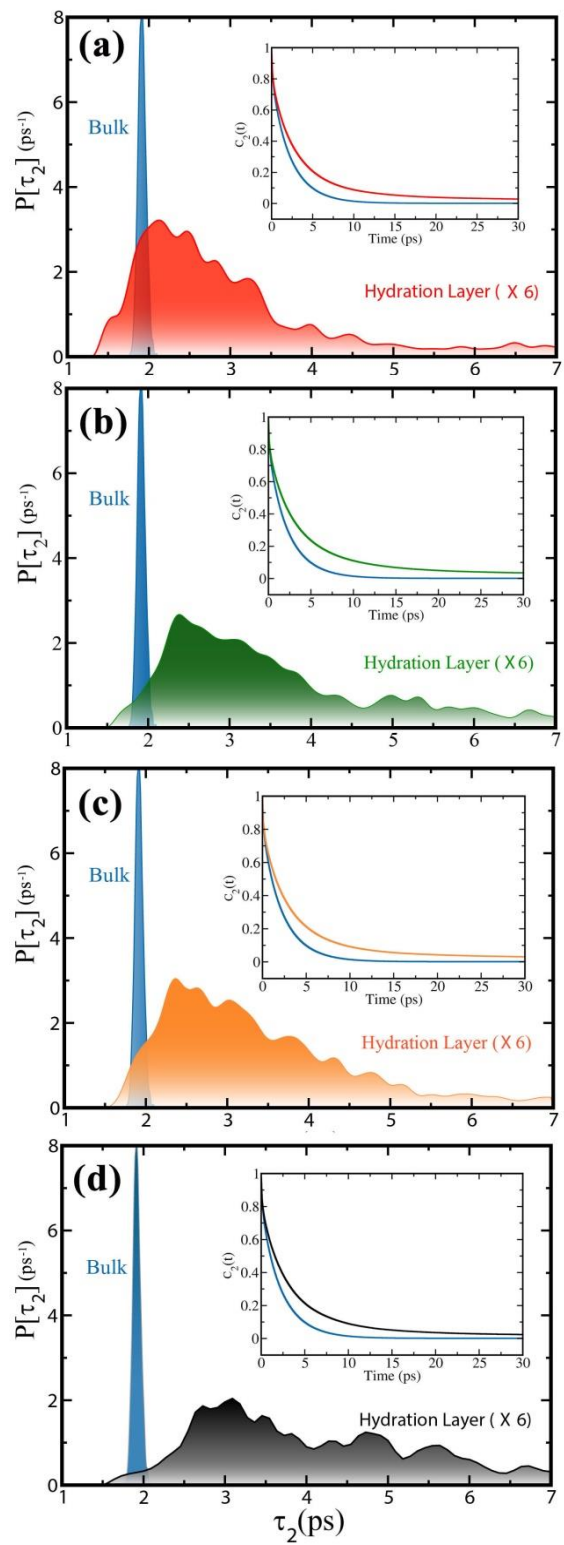

**Figure 3** Distribution of second rank rotational time constants of water molecules inside protein hydration layer for four proteins and bulk (blue). In the insets the normalised averaged rotational time correlation function is shown using same colour codes. (a) Lysozyme (b) Myoglobin (c) Monellin and (d) Protein-G.

**Table 2. Multi-exponential fitting parameters of the averaged and normalised second rank rotational relaxation of hydration layer water molecules. The slowest of the timescales (noted down in bold) was absent in bulk solvent.**

|  | $a_1$ | $\tau_1$(ps) | $a_2$ | $\tau_2$(ps) | $a_3$ | $\tau_3$(ps) | $<\tau>$(ps) | Average retardation |
|---|---|---|---|---|---|---|---|---|
| **Lysozyme** | 0.23 | 0.16 | 0.70 | 3.17 | **0.07** | **33.67** | 4.59 | 2.43 |
| **Myoglobin** | 0.25 | 0.18 | 0.67 | 3.55 | **0.08** | **37.56** | 5.81 | 3.07 |
| **Monellin** | 0.21 | 0.11 | 0.68 | 2.93 | **0.11** | **21.70** | 4.40 | 2.33 |
| **Protein-G** | 0.24 | 0.17 | 0.70 | 3.40 | **0.06** | **35.39** | 4.54 | 2.40 |
| **Bulk Water** | 0.22 | 0.13 | 0.78 | 2.39 | --- | --- | 1.89 | 1.00 |

**Figure 2** and **Figure 3** depict that the distributions in protein hydration layers are broad and it is a trademark of dynamic heterogeneity. More interestingly, there are few water molecules (~4-5% but varies from protein to protein) that relax faster than bulk water molecules along with a large fraction of slowly rotating water molecules. The faster rotating water molecules inside the hydration layer are proved to be those which are hydrogen bonded to protein backbone[54]. *The average rotational retardation factors are ~2.5-3.0 as compared to bulk.* This retardation factor has also been observed by NMR[29] and recent 2D-IR experiments[55]. If we look at the components of the relaxation, there is an extra timescale of amplitude ~18-25 % in the range of ~38-42 ps for Lysozyme, Myoglobin and Protein-G; and ~28 ps for Monellin is obtained which is absent in the case of bulk relaxation. The extra slow component arises presumably due to the long lived and strong hydrogen bonds that water forms with the charged residues (like Arg, Lys, Asp etc.) on the protein surface. Because of this kind of broad distribution PHL always shows heterogeneous dynamical responses. On the other hand, experimental techniques like NMR or 2D-IR are sensitive towards slow and ultrafast dynamics respectively. Moreover they provide only the average picture and not the microscopic details. Though this kind of detailed distributions cannot be obtained experimentally, MHRD technique claims to be successful at measuring the width of the distribution[29]. The broad spectrum of rotational relaxation pattern is responsible for the heterogeneous solvation dynamics throughout the PHL[33, 56].

## 3.2 Distribution of total dipole moment of hydration layer

Apart from the widely variant dynamical features of hydration layer and bulk water discussed in the previous section, some thermodynamic response functions are also quite efficient in discriminating between PHL and bulk. One of such properties is the effective dielectric constant of the shell which is a response function of total dipole moment fluctuation. The magnitude of total dipole moment ($M_T$) of a particular domain is given by **Equation(3)**.

$$M_T = \sqrt{\sum_j \left( \sum_{i=1}^{N} \mu_j^i \right)^2} \tag{3}$$

Here, *'i'* is the running index denoting water molecules and *'j'* is the index for Cartesian vector components (x,y,z) of dipole moment. This reflects a collective orientation of the water molecules. In presence of a huge and constant dipole moment arising from protein (see **Figure 1**), these orientations of water molecules are tremendously perturbed as compared to bulk, causing significant reduction in fluctuation of total dipole moment of that region.

**Figure 4** supports the foregoing discussion. It shows the distributions of total dipole moment fluctuation of the first hydration layer of four proteins compared to that of the bulk. Because of the decrease in fluctuation of total dipole moment in hydration layer as discussed above, a considerable narrowing of the distribution is observed. This distribution of bulk is obtained by constructing an analogous shell in bulk water maintaining same volume and shape of PHL.

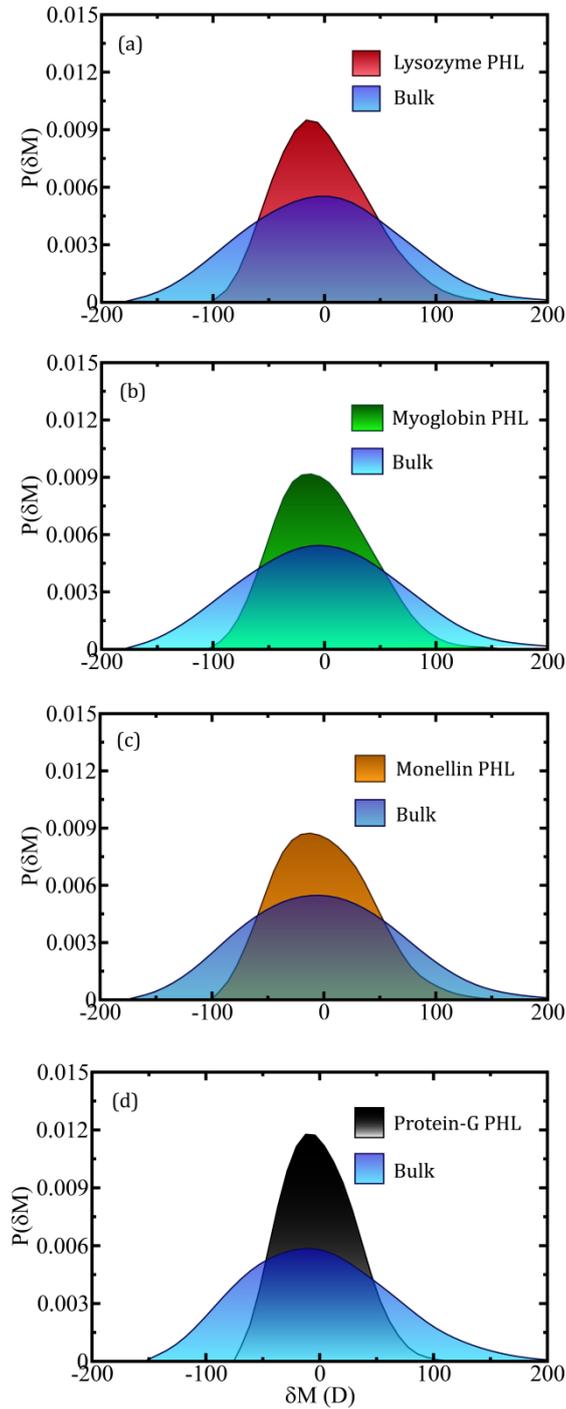

**Figure 4. Comparison of distributions of total dipole moment fluctuations in first hydration layer of four proteins and that of bulk (blue). (a) Lysozyme (red) (b) Myoglobin (green) (c) Monellin (orange) (d) Protein-G (black). The width of distribution in case of hydration layer becomes almost half as compared to the bulk.**

For this narrow distribution inside PHL, the dielectric constant becomes lower than that of bulk; as also observed by Ghosh *et. al*[42]. We calculate this property using the well known expression in terms of total dipole moment fluctuation[57-59] as shown in **Equation (4)**.

$$\varepsilon = 1 + \frac{4\pi}{3Vk_BT}\left\langle \left(M_T - \langle M_T \rangle\right)^2 \right\rangle \tag{4}$$

An important point to be noted in this context is that the definition of an '*effective dielectric constant*' of hydration layer is valid only in the limiting condition that cross correlation coefficient [**Equation(10)**] between total dipole moment fluctuations of PHL and the same of the next layer should be low (~10% or so). An analytical description of the issue is given below. Considering the total dipole moment of water to be $\bar{M}_W$, we can write

$$\left\langle \delta\bar{M}_W^2 \right\rangle = \left\langle \sum_i \delta\bar{M}_W^{i\,2} \right\rangle + 2\left\langle \sum_i \sum_{j\neq i} \delta\bar{M}_W^i \delta\bar{M}_W^j \right\rangle \tag{5}$$

Where, $i$ and $j$ are indices denoting shell around protein. Hence the total dipole moment of water has one self-part, and a cross-part. Now, if the cross-term is negligible, we can rewrite **Equation(5)** as

$$\left\langle \delta\bar{M}_W^2 \right\rangle = \left\langle \sum_i \delta\bar{M}_W^{i\,2} \right\rangle = \sum_i \left\langle \delta\bar{M}_W^{i\,2} \right\rangle \tag{6}$$

Scaling **Equation(6)** with respect to volume, we obtain,

$$\frac{\left\langle \delta\bar{M}_W^2 \right\rangle}{V_W} = \sum_i \frac{\left\langle \delta\bar{M}_W^{i\,2} \right\rangle}{V_W} = \sum_i \left( \frac{\left\langle \delta\bar{M}_W^{i\,2} \right\rangle}{V_W^i} \right) \times \left( \frac{V_W^i}{V_W} \right) \tag{7}$$

Where, $V_W$ is the volume of the total water and $V_W^i$ is the volume of the $i^{\text{th}}$ water shell. Multiplying both sides of **Equation(7)** with the factor $\frac{4\pi}{3k_BT}$ and defining volume fraction of $i^{\text{th}}$ shell as $v_f^i = \frac{V_W^i}{V_W}$, we get

$$\frac{4\pi \left\langle \delta\bar{M}_W^2 \right\rangle}{3V_W k_B T} = \frac{4\pi}{3k_BT} \sum_i \left( \frac{\left\langle \delta\bar{M}_W^{i\,2} \right\rangle}{V_W^i} \right) v_f^i$$

$$\text{hence, } \varepsilon_W = \sum_i v_f^i \varepsilon_i \tag{8}$$

where, $\varepsilon_W$ is the dielectric constant for all the water molecules in the system and $\varepsilon_i$ is the effective dielectric constant of the $i^{th}$ hydration shell, which is defined by **Equation(9)**.

$$\varepsilon_i^{eff} = 1 + \frac{4\pi}{3V_W^i k_B T}\left\langle \left(M_W^i - \left\langle M_W^i \right\rangle\right)^2 \right\rangle \quad (9)$$

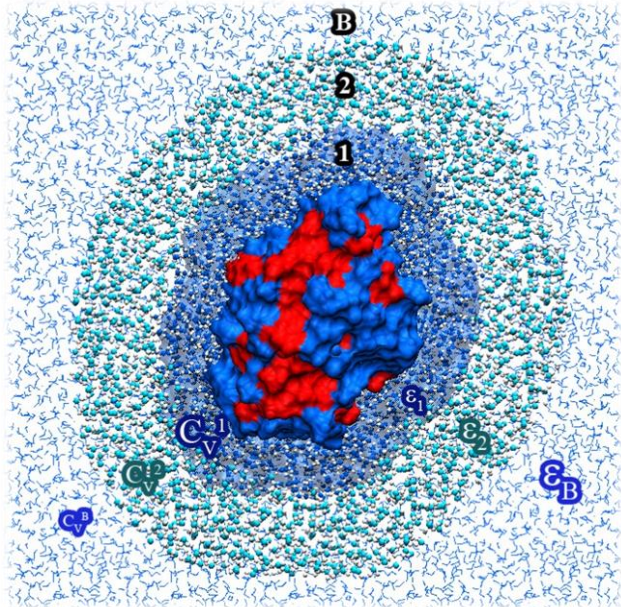

**Figure 5. Protein surrounded by water layers (cross-section); Layer-1 is the PHL. Effective dielectric constant of water shell increases and local specific heat decreases as we move away from protein; i.e., $\varepsilon_1 < \varepsilon_2 < \varepsilon_{Bulk}$, whereas $C_V^1 > C_V^2 > C_V^{Bulk}$**

The measure of smallness of the cross terms with respect to the self-terms are defined using the well-known expression of correlation coefficient ($\rho_{12}$) as shown in **Equation(10)**.

$$\rho_{12} = \frac{\text{cov}(M_1, M_2)}{\sqrt{\text{var}(M_1)\text{var}(M_2)}} = \frac{\left\langle \left(M_1 - \left\langle M_1 \right\rangle\right)\left(M_2 - \left\langle M_2 \right\rangle\right)\right\rangle}{\sqrt{\left\langle \left(M_1 - \left\langle M_1 \right\rangle\right)^2 \right\rangle \left\langle \left(M_2 - \left\langle M_2 \right\rangle\right)^2 \right\rangle}} \quad (10)$$

Here, angular brackets denote average over time. $M_1$ and $M_2$ are the total dipole moment of first and second layers respectively. The protein surface is generally rugged even for a globular protein. The width of the PHL and second layer are judicially chosen to be 1 nm and 2 nm respectively to avoid the effect of protein surface heterogeneity so that the cross correlation becomes negligible across the layers and there is not much discrepancy in the volume calculation in case we consider the shell to be spherical. To get the volume of hydration layer we have used the specific volume of water at 300 K along with the number density calculated from our MD trajectories.

**Table 3. Effective dielectric constants of the protein hydration layer, second layer and bulk water in case of four protein water systems. The cross-correlation coefficients are also tabulated and found to be ~10% compared to self-term. Indices '1', '2' and 'B' signifies PHL, shell-2 and bulk respectively.**

|  | Lysozyme | Myoglobin | Monellin | Protein-G |
|---|---|---|---|---|
| $\varepsilon^{eff}$ (Shell-1) | 46.54 | 44.39 | 48.67 | 43.24 |
| $\varepsilon^{eff}$ (Shell-2) | 54.01 | 52.38 | 63.25 | 55.96 |
| $\varepsilon$ (Bulk) | 68.77 | | | |
| $\rho_{12}$ | 0.15 | 0.09 | 0.12 | 0.14 |

We also calculate the total moment-moment autocorrelation function and compare it with the same in the bulk. The autocorrelation relaxations are fitted bi-exponential forms for PHL and single-exponential for Bulk water. There is always a slower component of one order of magnitude higher in case of PHL compared to bulk[42]. The relaxation is generally slower because of the large dipole moment of the protein which itself prevents the surrounding water dipoles to relax rapidly. For the beta sheet rich proteins GB1 and Monellin, the average retardation factor is ~1.5 and for that of alpha helix rich proteins Lysozyme and Myoglobin it is ~2.5. The fitting parameters are summarized in **Table 4** and the plots are shown in **Figure 5**.

**Table 4. Multi-exponential fitting parameters for <M(0)M(t)> of PHL and bulk solvent. There exist a slower component in case of PHL which is absent in bulk.**

|  | $a_1$ | $\tau_1$ (ps) | $a_2$ | $\tau_2$ (ps) | $<\tau>$ (ps) | Average Retardation |
|---|---|---|---|---|---|---|
| **Lysozyme** | 0.86 | 9.26 | 0.14 | 132.2 | 26.47 | 2.69 |
| **Myoglobin** | 0.85 | 8.64 | 0.15 | 120.05 | 25.35 | 2.58 |
| **Monellin** | 0.91 | 9.35 | 0.09 | 72.59 | 15.04 | 1.53 |
| **Protein-G** | 0.90 | 9.07 | 0.10 | 81.87 | 16.35 | 1.66 |
| **Bulk Solvent** | 1.00 | 9.81 | --- | --- | 9.81 | 1.00 |

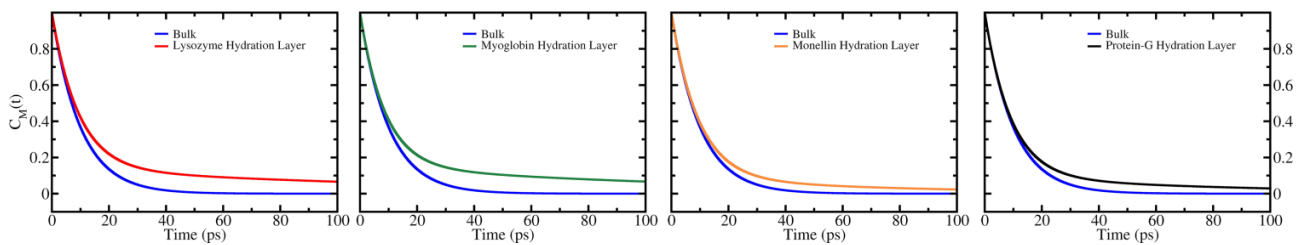

**Figure 6. Plots of total moment-moment autocorrelation function for protein hydration layer and bulk. The same of that of bulk solvent is shown in blue and for the proteins pervious colour codes are retained. Lysozyme (red), Myoglobin (green), Monellin (orange) and Protein-G (black).**

### 3.3 Distribution of water self-interaction energy of hydration layer

Besides dipole moment, total self-interaction energy (Coulomb and Lennard-Jones) distribution of PHL and bulk water molecules is also widely different. For bulk, it is sharp and narrow whereas in case of PHL, it is generally wider (**Figure 7**). For bulk the FWHM (Full Width at Half Maximum) is ~400 $k_B T$, whereas, the same for Lysozyme, Myoglobin, Monellin and Protein-G are ~1060, ~860, ~690 and ~1030 $k_B T$ respectively.

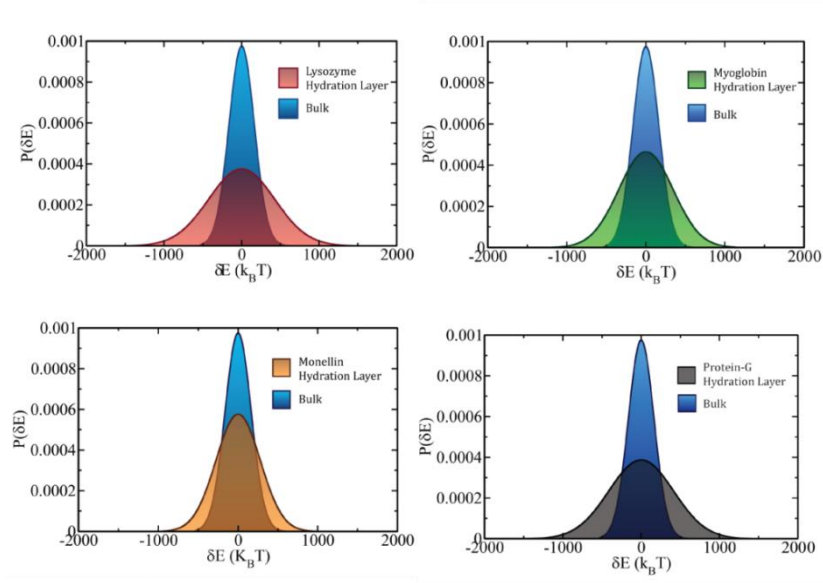

**Figure 7.** Distribution of total self-interaction energy of protein hydration layer compared with bulk (blue). For every protein-water system the distribution is broader than bulk. This implies a larger specific heat of the hydration layer sub-ensemble as in NVT ensemble the width of the distribution is proportional to the specific heat at constant volume.

Energy fluctuation is manifested in the form of the static response function specific heat[31] ($C_V$) given by **Equation(11)**.

$$C_V = \frac{1}{k_B T^2} \left\langle \left(E - \langle E \rangle\right)^2 \right\rangle \tag{11}$$

This energy fluctuation can have two contributions arising from potential energy and kinetic energy. For PHL, the potential or interaction energy term has two parts, one self-term and the other cross-term. Hence variance of potential energy for PHL can be expressed using **Equation(12)**.

$$\left\langle \left(\delta E^1\right)^2 \right\rangle = \left\langle \left(\sum_i \delta E_i\right)^2 \right\rangle + \left\langle \left(\sum_{i,j} \delta E_{i,j}\right)^2 \right\rangle + 2 \left\langle \left(\sum_i \delta E_i\right)\left(\sum_{i,j} \delta E_{i,j}\right) \right\rangle \tag{12}$$

$\sum_i \delta E_i$ is the self interaction energy among water molecules in PHL whereas $\sum_{i,j} \delta E_{i,j}$ is the cross interaction energy between molecules in PHL and rest of the system. Hence in the limiting condition that self-interaction is much greater than cross-interaction, we have

$$\lim_{\frac{\sum_{i,j}\delta E_{i,j}}{\sum_i \delta E_i} \to 0} \left\langle \left(\delta E^1\right)^2 \right\rangle = \left\langle \left(\sum_i \delta E_i\right)^2 \right\rangle \qquad (13)$$

This allows us to define a local specific heat of the PHL having self-energy contribution only, following the definition in **Equation(11)**.

We may derive specific heat like quantity for kinetic energy contribution as well, since this includes the individual molecules themselves. These results are tabulated in **Table 5**. It is observed that the specific heat values of PHL is more than twice of that of bulk values for both potential and kinetic energy contributions. The sum of the two gives the total effective heat capacity of different shells around the protein. In all the cases specific heat is found to be greater that twice that of bulk water.

**Table 5. Effective specific heat of PHL and shell-2 of four proteins compared to that of bulk. Table contains data for both potential and kinetic energy contributions. $C_V$ values are in cal K$^{-1}$ g$^{-1}$ unit.**

| Contribution | Shell # | Lysozyme | Myoglobin | Monellin | Protein-G | Bulk |
|---|---|---|---|---|---|---|
| Potential Energy | 1 | 1.89 | 1.79 | 1.69 | 1.74 | 0.74 |
| Potential Energy | 2 | 1.63 | 1.34 | 1.03 | 1.58 | |
| Kinetic Energy | 1 | 0.62 | 0.75 | 0.65 | 0.56 | 0.32 |
| Kinetic Energy | 2 | 0.77 | 0.90 | 0.45 | 0.71 | |
| Effective Specific heat | 1 | 2.51 | 2.54 | 2.34 | 2.30 | 1.06 |
| Effective Specific heat | 2 | 2.40 | 2.24 | 1.48 | 2.29 | |

A greater value of specific heat points towards a greater fluctuation in energy. Analogous to the case of dipole moment, energy of aqueous system is also highly perturbed by the presence of a large biomolecule like protein. The side chains of protein residues undergo continuous ceaseless conformational fluctuations which generate random kicks on the nearby water molecules. This results in an increased energy fluctuation in the hydration layer water molecules. Consequently, the specific heat of the layer also increases.

This increased specific heat is an indication of increased resistance towards temperature change of water molecules inside PHL because in NVT ensemble the specific heat at constant volume is proportional to the energy fluctuation [**Equation(11)**]. So it would be twice or thrice as difficult to change the local temperature of the PHL as it is in bulk. As the function of a particular protein is sensitive to the local temperature of the surroundings, PHL plays a huge role to provide that environment acting like a shield.

### 3.4 Translation time of water molecules inside hydration layer

We first define the residence time of a single water molecule as the time spent inside the ~1nm shell (chosen as the width of hydration layer) from the surface of a protein. We also compare it with the residence time of bulk water by concentrating on a similar sized shell which equals the PHL in volume, but without having the protein inside. We find that the *mean residence time* is within ~90-100 ps when the protein is present but reduces to only ~30-40 ps in the absence of the protein. From there we calculate the time required for a water molecule residing inside the PHL to translate by the same distance as its LJ diameter ($\sigma$=0.316 nm for SPC/E water model). We find that in bulk solvent this value averaged around 3.3 ps but in case of PHL we again see a broad distribution varying from protein to protein. All of the distributions have a distinct long tail extending up to ~30 ps − 40 ps. The distributions are given in **Figure 8**.

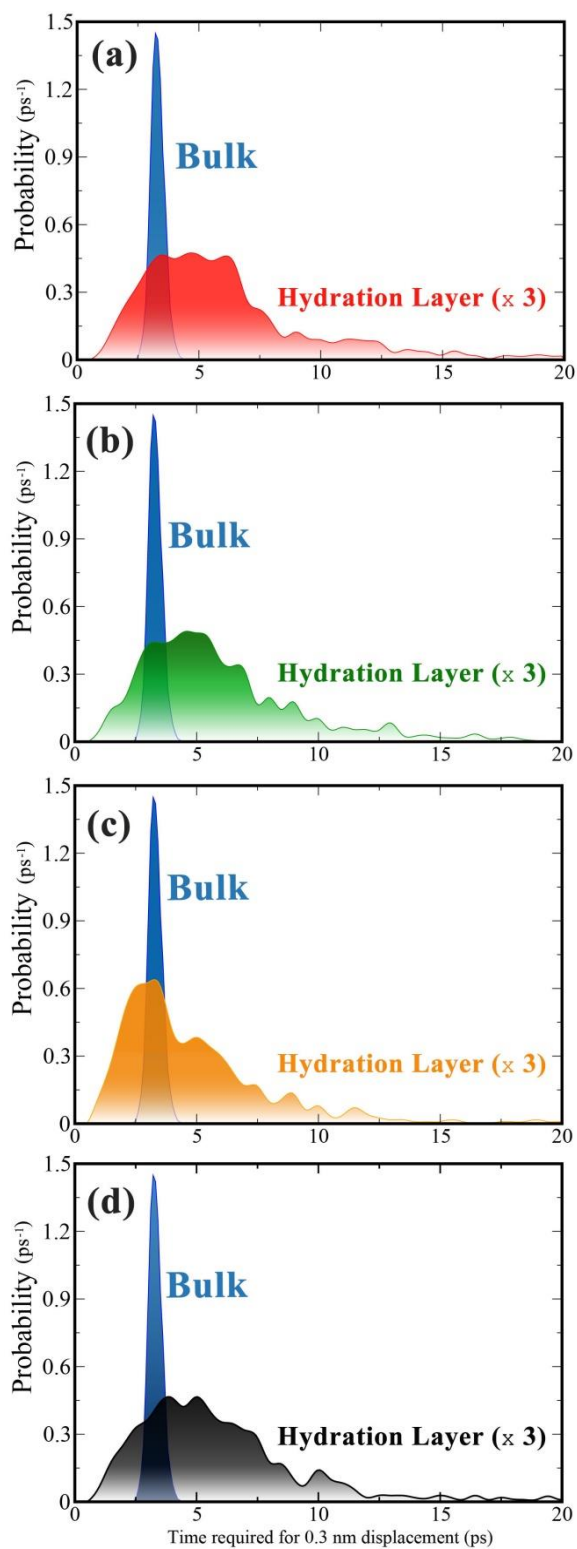

**Figure 8.** Broad distribution of the time required to get displaced equal to one molecular diameter for water molecules in bulk (shown in blue) and inside protein hydration layer ranging from 1ps to 20ps for (a) Lysozyme (red) (b) Myoglobin (green) (c) Monellin (orange) and (d) Protein-G (black). Noticeably there exist some water molecules which travel faster than bulk.

There are water molecules which translate faster than the molecules in the bulk along with the slower and bulk like ones. If we choose the average value of the sharp bulk distribution as the boundary to call a water molecule 'fast' or 'slow' we end up with the following numbers tabulated in **Table 6**. The faster translating ones are near hydrophobic regions facing a repulsive potential. The faster movement also arises from the 'kicking motion' produced by long and extended amino acid side-chains such as Arginine, Lysine etc. Because of the low rotational barrier[60], incessant side-chain conformation fluctuations introduce a constant perturbation to the hydration layer which in turn increases the energy content of the same. This is manifested in the high specific heat of hydration layer (see **Section 3.2**).

**Table 6. Fraction of fast and slow translating water molecules inside protein hydration layer of four different protein-water systems**

|  | % of translationally fast water | % of translationally slow water | Average time taken to translate by $\sigma$(ps) | Average retardation compared to bulk |
|---|---|---|---|---|
| **Lysozyme** | 23 | 77 | 6.9 | 2.09 |
| **Myoglobin** | 18 | 82 | 6.6 | 2.00 |
| **Monellin** | 35 | 65 | 5.3 | 1.61 |
| **Protein-G** | 27 | 73 | 6.0 | 1.82 |
| **Bulk Water** | --- | --- | 3.3 | 1.00 |

In order to quantify the obtained residence times with the help of a suitable time correlation function, we define s(t) [see **Equation (14)**], which is a measure of the lifetime of a water molecule inside the hydration layer. It is defined as,

$$s(t) = \frac{\langle h(0)h(t) \rangle}{\langle h(0)h(0) \rangle} \quad (14)$$

Here, 'h(t)' is a *heaviside step function*[61] at time '*t*' that describes the 'in-or-out' state of a water molecule. It takes up a value of '1' when the water molecule is inside the PHL and '0' otherwise. In addition to that we use an '*overlook period*' of 2 ps (which is small compared to the mean residence time inside the PHL). If a particular water molecule, located at the imaginary boundary of first shell and second shell, leaves the PHL for a duration which is less than the overlook period we consider that to be continuously inside the PHL. Once it is outside PHL for more than 2 ps, we consider that to be a '0' state from that time forever. This allows us to treat those water molecules which cannot get stabilised outside PHL and comes back again to the first shell. We calculate s(t) for each individual water molecules inside PHL and take an average over the molecules. The resultant time correlation functions are fitted to a multi-exponential (**Table 7, Figure 9**) and by integrating over time we extract the mean lifetime of water molecules inside PHL.

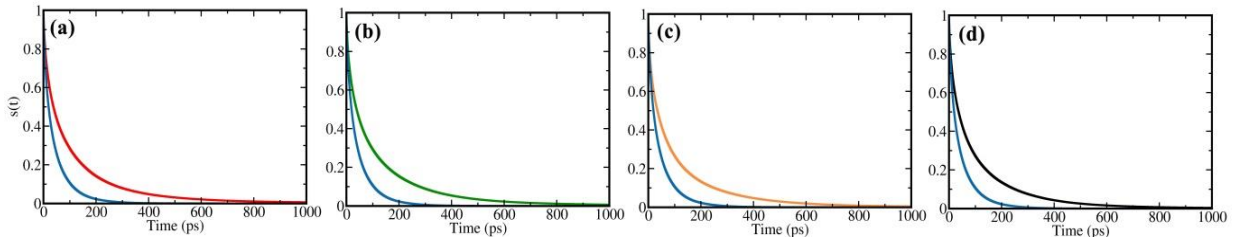

**Figure 9. Residence time correlation using heaviside step function formalism of PHL compared with the same in bulk water (blue). (a)Lysozyme (red) (b)Myoglobin (green) (c)Monellin (orange) and (d)Protein-G (black). The average time constant shows a ~2.5 times slowdown for PHL water molecules.**

**Table 7. Multi-exponential fitting parameters of the normalised residence time correlation function, s(t) using heaviside step function formalism for hydration layer water molecules and for bulk water.**

|  | $a_1$ | $\tau_1$(ps) | $a_2$ | $\tau_2$(ps) | $a_3$ | $\tau_3$(ps) | $<\tau>$(ps) | Average retardation |
|---|---|---|---|---|---|---|---|---|
| **Lysozyme** | 0.29 | 9.2 | 0.47 | 74.5 | 0.24 | **244.3** | 96.32 | 2.46 |
| **Myoglobin** | 0.25 | 7.5 | 0.41 | 58.8 | 0.34 | **219.6** | 100.6 | 2.57 |
| **Monellin** | 0.24 | 5.2 | 0.45 | 53.1 | 0.31 | **213.6** | 91.36 | 2.33 |
| **Protein-G** | 0.21 | 5.9 | 0.41 | 47.2 | 0.38 | **185.6** | 91.12 | 2.32 |
| **Bulk Water** | 0.17 | 3.4 | 0.36 | 21.4 | 0.47 | 65.7 | 39.16 | 1.00 |

From **Table 7** it is clear that the average retardations (defined as $<\tau>_{hyd}/<\tau>_{bulk}$) are ~2.3-2.6 compared to bulk. But the measure cannot promulgate the existing broad distribution which is the primary reason for uniqueness of PHL. Moreover, there exist one such timescale which is of one order of magnitude higher than that of bulk. But again the average value cannot capture this. Due to the presence of this heterogeneity arise the unique properties of PHL along with the site dependant local responses.

### 3.4 Heterogeneous solvation dynamics inside hydration layer

Because of the multitude of rotational and translational timescales inside PHL, the dynamics of solvation becomes a site dependent phenomenon throughout the hydration layer. In section 3.1 we have shown that there are a few water molecules that are rotating faster than bulk water molecules, although the majority of water molecules in the hydration layer are slower. In order to explore this aspect, we put virtual probes at different sites of the PHL. A virtual probe is a spherical point positive charge with 0.5 Å radius which is fixed with respect to an atom on the protein surface. We have used four spheres situated at different locations inside PHL for each protein to probe the dynamical response of different sites. The interaction energies are taken to be the sum of coulomb and LJ interactions[62]. Linear response theory[63] is applied on each energy trajectory to find out the solvation time correlation function[62, 64-67] [Equation (15)] and the timescales are obtained using a multi-exponential fitting equation with a Gaussian component to take care of the initial sub ~100 fs ultrafast decay[68] (see

Table 8 for details).

$$C(t) = \frac{<\delta E_{solv}(0)\delta E_{solv}(t)>_{gr}}{<\delta E_{solv}(0)^2>_{gr}} \quad (15)$$

Here, $\delta E_{solv}(t)$ is the fluctuation given by; $\delta E_{solv}(t) = E_{solv}(t) - <E>$. The subscript '$gr$' indicates averaging over ground state only[41].

The time constant of solvation of a bare ion in water is extremely fast[64, 65]. This can be partly (not fully) realised with the help of **Equation(16)**.

$$\tau_L = \left(\frac{\varepsilon_\infty}{\varepsilon_0}\right)\tau_D \quad (16)$$

Debye relaxation time $\tau_D$ is 8.3 ps for water. $\varepsilon_\infty$ and $\varepsilon_0$ are the infinite frequency and static dielectric constants for water. $\varepsilon_0$ is ~78 and $\varepsilon_\infty$ is ~5. Solvation relaxation time for an ion would then be ~500 fs is water. Now the value for $\varepsilon_0$ decreases as we move closer towards the protein surface[42]. This results in a slower solvation. But there are other governing factors as well, such as the inertial component and the heterogeneity of time scales. Because of the broad distributions of dynamical quantities inside PHL, different sites measure responses in a different manner when it comes to a partly local probe like solvation.

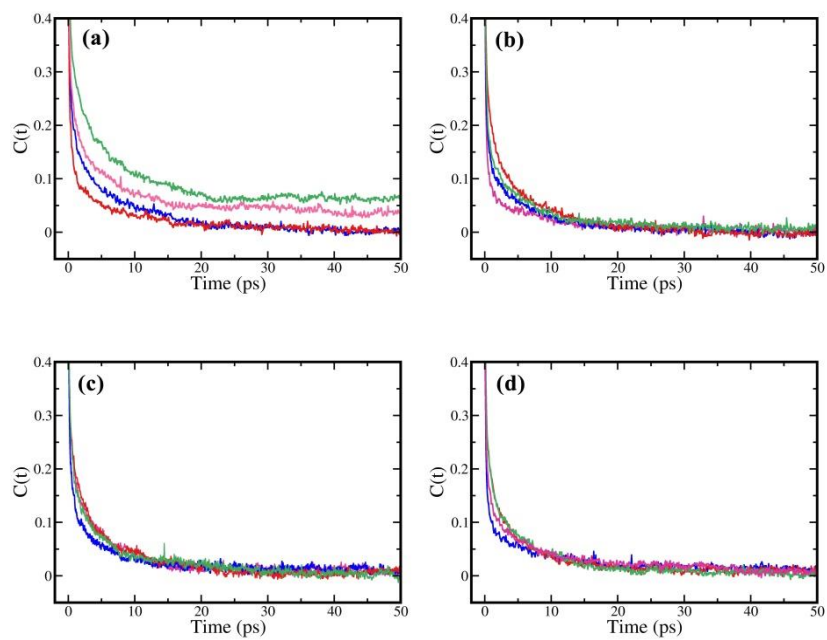

**Figure 10. Normalized total energy correlation plots for several virtual probes situated at different sites in the protein hydration layer of (a) Lysozyme, (b) Myoglobin, (c) Monellin and (d) Protein-G. The plots are shown only from C(t)=0.4 as the initial ~60-70% decay is ultrafast and ubiquitous. This difference in timescales shows the dynamical heterogeneity present inside the PHL.**

As expected, different locations show different timescales of solvation (**Figure 10**) though the average time constants are close to each other. As pointed out in earlier studies, solvation becomes slow near charged side-chains, not only due to slow water molecules but also because of the contribution of the charged/polar amino acid side chains[33]. The regions which contain fast rotating water molecules generally have faster solvation. These regions are near the backbone of protein and near hydrophobic groups.

However, as in the case of NMR, TDFSS also suffers from being able to measure only an average property except that one can use the location of the probe to get more insight into the distribution of relaxation times.

**Table 8. Multi-exponential fitting parameters for the solvation correlation function of virtual probes situated at different positions inside the protein hydration layer. The probe is placed**

within ~2-3 Å from a particular residue. The parameters are obtained after fitting the obtained normalised correlation to $C(t) = a_g e^{-\left(t/\tau_g\right)^2} + \sum_{i=1}^{n} a_i e^{-\left(t/\tau_i\right)}$.

| Protein | Probe Location | $a_g, \tau_g$ (ps) | $a_1, \tau_1$ (ps) | $a_2, \tau_2$ (ps) | $<\tau>$ (ps) |
|---|---|---|---|---|---|
| **Lysozyme** | near Trp-123 | 0.61, 0.077 | 0.25, 0.66 | 0.14, 9.42 | 1.53 |
| | near Trp-63 | 0.69, 0.074 | 0.23, 0.61 | 0.08, 12.22 | 1.16 |
| | near Trp-111 | 0.63, 0.095 | 0.28, 3.85 | 0.09, 107.53 | 10.81 |
| | near Trp-28 | 0.69, 0.089 | 0.23, 2.42 | 0.08, 50.24 | 4.63 |
| **Myoglobin** | near Tyr-146 | 0.64, 0.073 | 0.24, 0.42 | 0.12, 7.39 | 1.03 |
| | near His-12 | 0.59, 0.079 | 0.25, 0.72 | 0.16, 7.42 | 1.41 |
| | near His-81 | 0.62, 0.074 | 0.27, 0.59 | 0.11, 10.02 | 1.30 |
| | near His-113 | 0.70, 0.072 | 0.24, 0.41 | 0.06, 10.59 | 0.78 |
| **Monellin** | near Tyr-62 | 0.65, 0.047 | 0.28, 0.76 | 0.07, 16.16 | 1.37 |
| | near Tyr-78 | 0.53, 0.048 | 0.31, 0.58 | 0.16, 7.74 | 1.44 |
| | near Tyr-46 | 0.55, 0.052 | 0.33, 0.65 | 0.12, 10.21 | 1.46 |
| | near Tyr-28 | 0.50, 0.050 | 0.35, 0.49 | 0.15, 8.24 | 1.43 |
| **Protein-G** | near Val-29 | 0.74, 0.079 | 0.19, 0.96 | 0.07, 20.89 | 1.69 |
| | near Tyr-33 | 0.68, 0.085 | 0.24, 1.48 | 0.08, 13.66 | 1.50 |
| | near Tyr-3 | 0.59, 0.078 | 0.27, 0.64 | 0.14, 8.44 | 1.39 |
| | near Val-54 | 0.73, 0.082 | 0.19, 1.11 | 0.08, 18.51 | 1.74 |

## 4. CONCLUSION

As discussed extensively in the context of single molecule spectroscopy[69], the measured time correlation function is an ensemble averaged property. Just like we observed often in single molecule spectroscopy and also in super cooled liquids[29, 70], two different distributions can provide a similar time correlation function. It is thus possible to reach an erroneous conclusions if we base them on the ensemble average properties alone. The average can be a poor measure of reality.

The main results of the present work can be summarized as follows. Distributions of calculated residence times and rotational relaxation times in the hydration layer for four different protein-water systems are unusually broad. The distributions are Gaussian in bulk but transform to 'log-normal' in case of hydration layer. Note that, log-normality is abundant in nature[71]. We can justify the deviation from Gaussian to log-normality by assuming that the relaxation times scales by a multiplicative factor of $e^{-\beta \Delta E_i}$ with respect to bulk ($\Delta E_i = E_i^{PHL} - E_i^{Bulk}$, '$i$' is water index). Inside the PHL, if a particular water molecule becomes more stable than that of bulk, $\Delta E_i$ becomes negative. Hence, the factor takes up a positive value. As a result elongation of relaxation times occurs. On the contraty, there exists a fraction of water molecules which becomes less stable in PHL. For those, faster relaxation occurs (see Supporting Information for fits on Figure 8).

Physically, this arises from the simultaneous presence of (a) peptide backbones that form weak hydrogen bonds, (b) hydrophobic amino acid side chains that form no hydrogen bond and (c) charged polar groups that form strong hydrogen bond with the surrounding water molecules. This broad distribution is not reflected in the average time constant which is just about 2-3 times larger than that of bulk water[29, 34, 55] (**Tables 4** and **6**). In particular, the average leads to the erroneous conclusion that water in the hydration layer is bulk-like. Nevertheless, the mathematical description of log-normality is elusive and still deserves proper quantification.

Protein hydration layer is unique because the water molecules encounter highly heterogeneous surface with respect to structure and electrostatics. The water molecules that are hydrogen bonded to the peptide back-bone are known to rotate and translate faster than those bonded to charge groups like arginine or glutamate or aspartate[54]. In addition, the solvent exposed hydrophobic amino acid side chains offer no specific resistance to rotation of water molecules. Since different proteins could have substantially different sequence of amino acid residues, water dynamics could be quite sensitive to the specific nature of a particular protein. However, some aspects are conserved. As the peptide backbone should always be present, a part of relaxation of the hydration layer should always be faster than the bulk. The same goes for the water molecules near the hydrophobic residues. Therefore, one should be particularly concerned about the sensitivity to the charged amino acid groups. Only these groups can give rise to slower than bulk decay. Experiments measure the ensemble averaged retardation factors which are obtained in our present study as well but cannot bring out the true characterisation[29, 55]. The universal and defining characteristics of hydration layer would then be a broad distribution of relaxation times, with relaxation times substantially shorter than bulk to a range substantially higher than bulk.

On top of that due to its low dielectric constant and high specific heat it offers a unique property. Due to the low dielectric constant, the PHL cannot screen the interactions proteins do with ligands[72] (substrates, small molecules, drugs etc.) which are the most important part before any protein action such as enzyme kinetics[73] or aggregations[74, 75]. The high value of specific heat provides a protective environment around the protein which is more resistant to the temperature change that water molecules at the far. This helps the protein to function properly. We also discuss some aspects in favour of the uniqueness of the hydration layer. There are many other structural and dynamical properties (radial distribution, tetrahedral order parameter, dynamics structure factor, $\chi_4(t)$ etc.) that would serve equally good in this purpose. This can be extended to other biological macromolecules like DNA, as well given that their surface it heterogeneous.

The wide distribution of relaxation times could have the following important experimental ramifications. (i) The solvation dynamics could be highly non-exponential, as discussed. We have observed elsewhere that solvation dynamics can observe amino acid side chain motions and can be sensitive to slower than average dynamics.[1, 12, 32, 33] One needs to untangle the observed dynamics to obtain water contribution. (ii) NMR experiments that isotope label peptide group atoms can preferentially probe the faster motions of, weakly hydrogen bonded water molecules[28, 29, 76], (iii) 2D-IR experiments may also preferentially observe faster water molecules as these experiments also use isotope labelling of peptide group atoms[29]. Thus, results of both 2D-IR and NMR can be biased towards molecules exhibiting faster than average dynamics. Therefore, one needs to employ all the available techniques to understand and explore the wide distribution reported here, for the first time.

## 5. ACKNOWLEDGEMENTS

We are thankful to Dr. Rajib Biswas for many useful discussions and technical help. We thank the Department of Science and Technology (DST, India) for partial support of this work. B. Bagchi thanks sir J C Bose fellowship for partial support. S. Mondal thanks UGC, India for providing research fellowship and S. Mukherjee thanks DST, India for providing INSPIRE fellowship.

# Supporting information

## S1. Calculation of orientational relaxation times of PHL water

The orientational correlations calculated in **Section 3.1** correspond to first and second rank Legendre polynomials as given by **Equations (1)** and **(2)**. We followed the following steps to obtain the histograms shown in **Figures (2)** and **(3)** in the main text:

$$C_1(t) = \langle P_1(\hat{\boldsymbol{\mu}}_0 \cdot \hat{\boldsymbol{\mu}}_t) \rangle; \text{ where } P_1(x) = x \tag{17}$$

$$C_2(t) = \langle P_2(\hat{\boldsymbol{\mu}}_0 \cdot \hat{\boldsymbol{\mu}}_t) \rangle; \text{ where } \mathbf{P}_2(x) = \tfrac{1}{2}(3x^2 - 1) \tag{18}$$

1. We calculate the distance of each water molecule from protein atoms at every time-step throughout the MD trajectory. We define a cut-off radius of 1 nm for the PHL and select those water molecules which reside within this distance. This gives us water molecules inside the PHL (~2300-2500 at an average).
2. Then we calculate the residence times of these water molecules in the PHL and create an output file which contains data of entry-time, exit-time and total residence time of these molecules in the PHL.
3. We provide this file as an input into the program which calculates $C_1(t)$ and $C_2(t)$. It reads the data of entry-time and exit-time and computes the said terms according to **Equations (1)** and **(2)** for water molecules residing more than 100 ps in the PHL. This gave us one time correlation function (TCF) for every molecule considered.
4. We then fit the normalized data in each file to a multi-exponential function (which gave the best fit according to minimum $\chi^2$ values), and computed average relaxation time $\langle \tau \rangle$ for each water molecule using **Equation (3)**

$$\langle \tau \rangle = \sum_i a_i' \tau_i \tag{19}$$

5. Thereafter, we generated a histogram of these relaxation times which led to **Figures (2) and (3)** in main text.

To calculate the average time correlation functions shown in insets of **Figures (2)** and **(3)**, we have first averaged over all the water molecules and fit the resultant data according to tri-exponential function. The fitting parameters are presented in **Tables (1)** and **(2)** in the main text.

Similar procedure has been followed for all the four proteins.

We present some exemplary plots of individual TCFs for the PHL of lysozyme below.

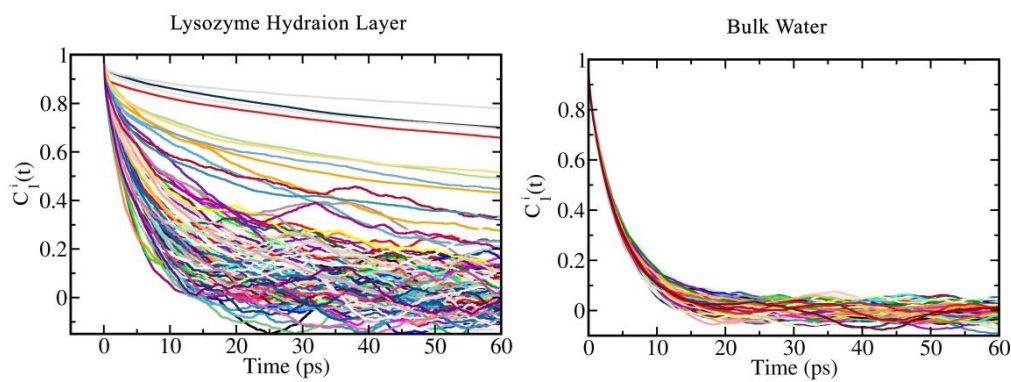

**Figure S1. Plots of first rank rotational time correlation functions of individual water molecules inside PHL (left) and for that of the bulk (right) in lysozyme-water system.**

Figure S1 shows individual TCFs for representative ~100 water molecules in the PHL and bulk. As clearly seen from the plots, some water molecules are rotationally rigid resulting is slowly decaying TCF and contributing to the long tail of the histogram shown in Fig. 2 and 3 of main text.

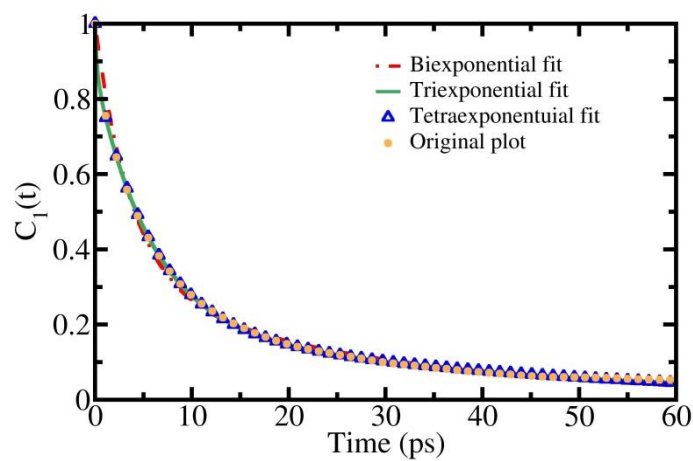

**Figure S2. Various multi-exponential fits to the particle averaged first rank orientational correlation function obtained for of Lysozyme-hydration layer.**

**Table S1.** $\chi^2$ values for various multi-exponential fits (Fig. S2) to describe the time correlation functions (for lysozyme-water system).

| Fitting Function | $a_1$ | $\tau_1$(ps) | $a_2$ | $\tau_2$(ps) | $a_3$ | $\tau_3$(ps) | $a_4$ | $\tau_4$(ps) | $\chi^2$ |
|---|---|---|---|---|---|---|---|---|---|
| Biexponential | 0.69 | 3.78 | 0.31 | 28.34 | -- | -- | -- | -- | 0.11371 |
| Triexponential | 0.13 | 0.21 | 0.66 | 5.63 | 0.21 | 38.65 | -- | -- | **0.00607** |
| Tetraexponential | 0.13 | 0.23 | 0.65 | 5.65 | 0.36 | 38.68 | 0.14 | 38.43 | 0.00621 |

We also provide the reason behind choosing a tri-exponential fit by evaluating $\chi^2$ values. As seen from Table S1, bi-exponential fit is not suitable whereas tetra-exponential functions produce almost the same result as tri-exponential. Same statistical treatment is done for other plots as well.

## S2. Calculation of translation times

In **Figure 8** (main text) we plot histograms of times taken by PHL water molecules to obtain a displacement equal to its molecular diameter ($\sigma$ = 0.316 nm for SPC/E water). PHL water has been selected according to **step 1** described in the previous **Section S1**. Then we calculate the displacements of each and every PHL water molecule at each time-step. The time (t) at which the difference between the final (time = t) and initial (time = 0) positions (displacement) becomes greater than $\sigma$, has been recorded for each molecule. **Figure 8** shows histograms of these translation times for PHL water of four proteins. The histograms for PHL have been multiplied by 3 for better comparison with bulk.

## S3. LOG-NORMAL NATURE OF THE DISTRIBUTIONS

We find that the nature of the distributions obtained for PHL is skew-symmetric with a long tail. On further investigation we observe that the distributions are log-normal in nature. This feature is universally present for every distribution. The mathematical form of a log-normal distribution is the following,

$$P(x) = \frac{1}{x\sigma\sqrt{2\pi}} e^{-\frac{1}{2\sigma^2}(\ln(x)-\mu)^2} \tag{20}$$

Where, μ is the mean and σ is the standard deviation. Here, for demonstration of this fact, we provide fitting parameters for the distributions in **Figure 8** of main text. We report the fitting parameters after transforming them from the log scale to the non-log scale in **Table S2**. The plots are given in **Figure S3**.

**Table S2. Fitting parameters for log-normal distribution of translational time distributions of PHL for four proteins. Bulk water fit is Gaussian in nature.**

| Proteins | Mean ($e^{\mu+\frac{\sigma^2}{2}}$) | Standard Deviation ($\{e^{\sigma^2+2\mu}(e^{\sigma^2}-1)\}^{1/2}$) |
|---|---|---|
| **Lysozyme** | 5.85 | 3.68 |
| **Myoglobin** | 5.75 | 3.39 |
| **Monellin** | 4.72 | 3.23 |
| **Protein-G** | 5.94 | 3.68 |
| **Bulk Water** | 3.28 | 0.27 |

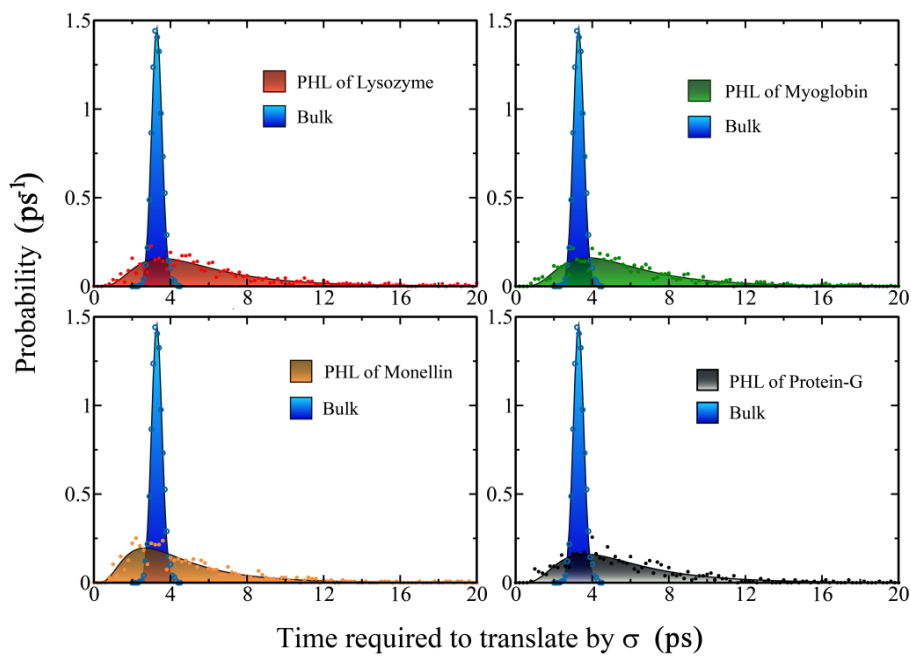

**Figure S3. Gaussian fit for bulk and Log-normal fits for protein hydration layer for four different proteins.**

Table S2 also supports the same story. At an average PHL is almost ~2 times slower than bulk. But the long tail in the distribution and some faster components depicts a different picture altogether. However, the proper mathematical investigation of the origin of log-normality is elusive and yet to be achieved.

**References:**

1. B. Bagchi, Proc. Natl. Acad. Sci. U.S.A. 113 (30), 8355-8357 (2016).
2. D. Laage, T. Elsaesser and J. T. Hynes, Chem. Rev. DOI: 10.1021/acs.chemrev.6b00765 (2017).
3. H. Frauenfelder, G. Chen, J. Berendzen, P. W. Fenimore, H. Jansson, B. H. McMahon, I. R. Stroe, J. Swenson and R. D. Young, Proc. Natl. Acad. Sci. U.S.A. 106 (13), 5129-5134 (2009).
4. B. Bagchi, Chem. Rev. 105 (9), 3197-3219 (2005).
5. D. Laage, G. Stirnemann, F. Sterpone, R. Rey and J. T. Hynes, Annu. Rev. Phys. Chem. 62, 395-416 (2011).
6. P. Ball, Proc. Natl. Acad. Sci. U.S.A., 201703781 (2017).
7. L. Zhang, L. Wang, Y.-T. Kao, W. Qiu, Y. Yang, O. Okobiah and D. Zhong, Proc. Natl. Acad. Sci. U.S.A. 104 (47), 18461-18466 (2007).
8. J. C. Rasaiah, S. Garde and G. Hummer, Annu. Rev. Phys. Chem. 59, 713-740 (2008).
9. W. Qiu, Y.-T. Kao, L. Zhang, Y. Yang, L. Wang, W. E. Stites, D. Zhong and A. H. Zewail, Proc. Natl. Acad. Sci. U.S.A. 103 (38), 13979-13984 (2006).
10. W. Qiu, L. Zhang, O. Okobiah, Y. Yang, L. Wang, D. Zhong and A. H. Zewail, J. Phys. Chem. B 110 (21), 10540-10549 (2006).
11. B. Bagchi, *Water in Biological and Chemical Processes: From Structure and Dynamics to Function*. (Cambridge University Press, 2013).
12. D. Zhong, S. K. Pal and A. H. Zewail, Chem. Phys. Lett. 503 (1), 1-11 (2011).
13. S. M. Bhattacharyya, Z.-G. Wang and A. H. Zewail, J. Phys. Chem. B 107 (47), 13218-13228 (2003).
14. N. Nandi and B. Bagchi, J. Phys. Chem. B 101 (50), 10954-10961 (1997).
15. R. Pethig, Annu. Rev. Phys. Chem. 43 (1), 177-205 (1992).
16. E. Grant, Bioelectromagnetics 3 (1), 17-24 (1982).
17. G. Otting, E. Liepinsh and K. Wuthrich, Science 254 (5034), 974-980 (1991).
18. K. Wüthrich, M. Billeter, P. Güntert, P. Luginbühl, R. Riek and G. Wider, Faraday Discuss. 103, 245-253 (1996).
19. D. Svergun, S. Richard, M. Koch, Z. Sayers, S. Kuprin and G. Zaccai, Proc. Natl. Acad. Sci. U.S.A. 95 (5), 2267-2272 (1998).
20. Y. Levy and J. N. Onuchic, Annu. Rev. Biophys. Biomol. Struct. 35, 389-415 (2006).
21. M. Chaplin, Nature Reviews Molecular Cell Biology 7 (11), 861-866 (2006).
22. P. Ball, CELLULAR AND MOLECULAR BIOLOGY-PARIS-WEGMANN- 47 (5), 717-720 (2001).
23. R. A. Robinson and R. H. Stokes, *Electrolyte solutions*. (Courier Corporation, 2002).
24. N. Nandi, K. Bhattacharyya and B. Bagchi, Chem. Rev. 100 (6), 2013-2046 (2000).
25. S. Mashimo, S. Kuwabara, S. Yagihara and K. Higasi, J. Phys. Chem. 91 (25), 6337-6338 (1987).
26. S. Mashimo, S. Kuwabara, S. Yagihara and K. Higasi, J. Chem. Phys. 90 (6), 3292-3294 (1989).
27. V. P. Denisov and B. Halle, Faraday Discuss. 103, 227-244 (1996).
28. B. Halle, Philos. Trans. R. Soc. London, Ser. B 359 (1448), 1207-1224 (2004).
29. C. Mattea, J. Qvist and B. Halle, Biophysical journal 95 (6), 2951-2963 (2008).
30. S. K. Pal, J. Peon, B. Bagchi and A. H. Zewail, J. Phys. Chem. B 106 (48), 12376-12395 (2002).
31. B. Bagchi, *Molecular relaxation in liquids*. (OUP USA, 2012).
32. T. Li, A. A. Hassanali, Y.-T. Kao, D. Zhong and S. J. Singer, J. Am. Chem. Soc. 129 (11), 3376-3382 (2007).
33. S. Mondal, S. Mukherjee and B. Bagchi, Chem. Phys. Lett. 683, 29(2017).
34. F. Sterpone, G. Stirnemann and D. Laage, J. Am. Chem. Soc. 134 (9), 4116-4119 (2012).
35. B. Halle and L. Nilsson, J. Phys. Chem. B 113 (24), 8210-8213 (2009).
36. K. Bhattacharyya, Acc. chem. Res. 36 (2), 95-101 (2003).

37. K. Bhattacharyya, Chem. Comm. (25), 2848-2857 (2008).
38. S. K. Pal, D. Mandal, D. Sukul, S. Sen and K. Bhattacharyya, J. Phys. Chem. B 105 (7), 1438-1441 (2001).
39. S. K. Pal, J. Peon and A. H. Zewail, Proc. Natl. Acad. Sci. U.S.A. 99 (24), 15297-15302 (2002).
40. D. Zhong, S. K. Pal, D. Zhang, S. I. Chan and A. H. Zewail, Proc. Natl. Acad. Sci. U.S.A. 99 (1), 13-18 (2002).
41. K. Furse and S. Corcelli, J. Phys. Chem. Lett. 1 (12), 1813-1820 (2010).
42. R. Ghosh, S. Banerjee, M. Hazra, S. Roy and B. Bagchi, J. Chem. Phys. 141 (22), 22D531 (2014).
43. J. B. Hubbard and P. G. Wolynes, J. Chem. Phys. 69 (3), 998-1006 (1978).
44. S. Ravichandran and B. Bagchi, International Reviews in Physical Chemistry 14 (2), 271-314 (1995).
45. M. Heinig and D. Frishman, Nucleic acids research 32 (suppl 2), W500-W502 (2004).
46. B. Hess, C. Kutzner, D. Van Der Spoel and E. Lindahl, J. Chem. Theo. Comp. 4 (3), 435-447 (2008).
47. W. Humphrey, A. Dalke and K. Schulten, Journal of molecular graphics 14 (1), 33-38 (1996).
48. W. L. Jorgensen and J. Tirado-Rives, J. Am. Chem. Soc. 110 (6), 1657-1666 (1988).
49. S. Kirkpatrick, C. D. Gelatt and M. P. Vecchi, science 220 (4598), 671-680 (1983).
50. G. Bussi, D. Donadio and M. Parrinello, J. Chem. Phys. 126 (1), 014101 (2007).
51. M. Parrinello and A. Rahman, Physical Review Letters 45 (14), 1196 (1980).
52. B. Hess, H. Bekker, H. J. Berendsen and J. G. Fraaije, J. Comput. Chem. 18 (12), 1463-1472 (1997).
53. T. Darden, D. York and L. Pedersen, J. Chem. Phys. 98 (12), 10089-10092 (1993).
54. B. Jana, S. Pal and B. Bagchi, Journal of Chemical Sciences 124 (1), 317-325 (2012).
55. J. T. King, E. J. Arthur, C. L. Brooks III and K. J. Kubarych, J. Phys. Chem. B 116 (19), 5604-5611 (2012).
56. Y. Qin, M. Jia, J. Yang, D. Wang, L. Wang, J. Xu and D. Zhong, J. Phys. Chem. Lett. 7 (20), 4171-4177 (2016).
57. B. Bagchi and A. Chandra, J. Chem. Phys. 90 (12), 7338-7345 (1989).
58. J. G. Kirkwood and J. B. Shumaker, Proc. Natl. Acad. Sci. U.S.A. 38 (10), 855-862 (1952).
59. M. Neumann, Mol. Phys. 50 (4), 841-858 (1983).
60. R. J. Smith, D. H. Williams and K. James, Journal of the Chemical Society, Chemical Communications (11), 682-683 (1989).
61. T. Von Karman and M. A. Biot, *Mathematical methods in engineering*. (McGraw Hill, 1940).
62. F. O. Raineri, H. Resat, B. C. Perng, F. Hirata and H. L. Friedman, J. Chem. Phys. 100 (2), 1477-1491 (1994).
63. P. Hänggi and H. Thomas, Physics Reports 88 (4), 207-319 (1982).
64. R. Jimenez, G. R. Fleming, P. Kumar and M. Maroncelli, Nature 369, 471-473 (1994).
65. M. Maroncelli, Journal of Molecular Liquids 57, 1-37 (1993).
66. M. Maroncelli, J. Chem. Phys. 94 (3), 2084-2103 (1991).
67. M. Maroncelli and G. R. Fleming, J. Chem. Phys. 89 (8), 5044-5069 (1988).
68. B. Bagchi and B. Jana, Chem. Soc. Rev. 39 (6), 1936-1954 (2010).
69. T. Plakhotnik, E. A. Donley and U. P. Wild, Annu. Rev. Phys. Chem. 48 (1), 181-212 (1997).
70. P. G. Debenedetti and F. H. Stillinger, Nature 410 (6825), 259-267 (2001).
71. E. Limpert, W. A. Stahel and M. Abbt, BioScience 51 (5), 341-352 (2001).
72. M. Wilchek, E. A. Bayer and O. Livnah, Immunology letters 103 (1), 27-32 (2006).
73. I. H. Segel, *Enzyme kinetics*. (Wiley, New York, 1975).
74. R. R. Kopito, Trends in cell biology 10 (12), 524-530 (2000).
75. C. A. Ross and M. A. Poirier, (2004).
76. K. Modig, E. Liepinsh, G. Otting and B. Halle, J. Am. Chem. Soc. 126 (1), 102-114 (2004).